\documentclass[aps,prl,twocolumn,superscriptaddress,groupedaddress,nofootinbib]{revtex4}  
\usepackage{graphicx}  
\usepackage{dcolumn}   
\usepackage{bm,amsmath}        
\usepackage{amssymb}   
\usepackage{color}

\newcommand{\beq}{\begin{equation}} 
\newcommand{\eeq}{\end{equation}} 
\newcommand{\bea}{\begin{eqnarray}} 
\newcommand{\eea}{\end{eqnarray}}

\begin{document}

\hspace{5.24in} 
\mbox{CERN-PH-TH-2013-165} 
\vspace{0.2in}
\hspace{5.2in} 
\mbox{FERMILAB-FN-0963-CD-T}

\title{The Matrix Element Method: Past, Present, and Future}
\author{James~S.~Gainer} \affiliation{Physics Department, University
  of Florida, Gainesville, FL 32611, USA}
\author{Joseph~Lykken} \affiliation{Theoretical Physics Department, Fermilab, Batavia, IL 60510, USA}
\author{Konstantin~T.~Matchev} \affiliation{Physics Department,
  University of Florida, Gainesville, FL 32611, USA}
\author{Stephen~Mrenna} \affiliation{SSE Group, Computing Division, Fermilab, Batavia, IL 60510, USA}
\author{Myeonghun~Park} \affiliation{Theory Division, Physics
  Department, CERN, CH-1211 Geneva 23, Switzerland}
\date{July 26, 2013}

\begin{abstract}
The increasing use of multivariate methods, and in particular the
Matrix Element Method (MEM), represents a revolution in
experimental particle physics.  With continued exponential growth
in computing capabilities, the use of sophisticated multivariate
methods-- already common-- will soon become ubiquitous and ultimately
almost compulsory.  
While the existence of sophisticated algorithms for disentangling
signal and background might naively suggest a diminished role for
theorists, the use of the MEM, with its inherent connection to the
calculation of differential cross sections will benefit from
collaboration between theorists and experimentalists.  In this white
paper, we will briefly describe the MEM and some of its recent uses,
note some current issues and potential resolutions, and speculate
about exciting future opportunities.

\end{abstract}

\maketitle

\textbf{Introduction--}
Multivariate methods~\cite{Bhat:2010zz} are widely used in
experimental particle physics; popular examples include boosted
decision trees (BDT) and neural nets, in addition to the Matrix
Element Method (MEM)~\cite{matrix}, which will be the subject
of our discussion here.  In the MEM, the likelihood for a given event,
with measured momenta $\textbf{p}_i^{\rm vis}$ in some
underlying model with parameters $\alpha$ is given by
\begin{equation}
\begin{aligned} 
&{\cal P}(\textbf{p}^{\rm vis}_i|\alpha) = \frac{1}{\sigma(\alpha)}
\sum_{k,l} \int dx_1 dx_2 \, \frac{f_k(x_1)f_l(x_2)}{2sx_1x_2} \\
&\qquad\qquad \times \biggl [ \prod_{j \in \text{inv.}} \int
 \frac{d^3 p_j}{(2\pi)^3 2E_j} \biggr] |{\cal M}_{kl}(p_i^{\rm vis},p_j;\alpha)|^2,
\end{aligned}
 \label{eq:mem1}
\end{equation}
where $f_k$ and $f_l$ are parton distribution functions, ${\cal
  M}_{kl}$ is the theoretical matrix element, and $\sigma_(\alpha)$ is
the (total) cross section after cuts and efficiencies.   If the process involves invisible
particles, such as neutrinos or neutralinos, then their momenta, $p_j$,
must be integrated over the appropriate phase space.

Transfer functions parameterizing the detector resolution, should also
be included and integrated over, as the matrix element is a function
of the actual, rather than the observed, particle momenta.
In the limit where all the quantities and functions in
Eq.~\ref{eq:mem1} are known with perfect accuracy, the quantity
calculated in this manner is the likelihood, and hence by the
Neyman-Pearson Lemma is an optimal test statistic~\cite{Neyman:Pearson}.
Hence, if it can be implemented, the MEM should be the most sensitive
analysis \emph{possible}-- we will note some caveats later in this
paper.

\textbf{Past and Present--}
The MEM has been used in studies of top properties at the
Tevatron~\cite{memtev}, as well as in B
physics~\cite{B physics}.  
A notable recent application has been in the study of $H \to ZZ^\ast
\to 4\ell$~\cite{memzz, memhzz nlo, jhu mem, mekd, geolocating}, where
the MEM, in the form of the ``MELA'' approach~\cite{jhu mem}, was used
in the Higgs discovery by CMS~\cite{MELA CMS}.  Currently CMS uses MELA
together with MEKD~\cite{mekd, MELA MEKD CMS}, another MEM package in
continuing studies of the properties of the Higgs.  ATLAS is using a
BDT for similar studies~\cite{ATLAS}.

The MEM has also been suggested for studying the Higgs in other
channels~\cite{mem higgs, MEM HiggsNLO} and for BSM energy frontier
physics~\cite{MEM BSM}.  A dedicated package, MadWeight, has been
developed for MEM studies~\cite{MadWeight}.

\textbf{Near Future--}
A practical challenge associated with the use of the MEM is that the
likelihood calculated is only an approximation of the true likelihood.
This situation arises because of (i) finite detector resolution (ii)
higher order corrections (iii) neglected information.  Therefore,
rather than taking e.g. the likelihood ratio calculated from the MEM
directly to have a given statistical significance, it may still be
necessary for an experiment to calculate the statistical significance
of a given result using e.g. pseudoexperiments, which naively can be
very expensive from the standpoint of computer time.

We note that the detector response to an event is generally
independent of the underlying parameters of the model which described
the hard process (e.g. masses or couplings of virtual particles
produced in the collision).
Thus one can often simply re-weight the events already generated in a
given pseudo-experiment to calculate the likelihood at {\em any} other
point in the parameter space~\cite{geolocating}.
This procedure eliminates the need to generate a separate event sample
for each point in parameter space and thus significantly speeds up the analysis. 
An analogous procedure could be employed for template-based analyses.

\textbf{Future--}
Moore's Law~\cite{Moore} predicts that the vast increases in computing
power that have characterized the past decades will continue.
Even today, technological advances such as the use of graphics
processing units~\cite{GPU} may significantly increase the computing
resources available for analyses.
If we
assume that this trend will continue, then ultimately we will have the
computing power to perform analyses that would be wildly impractical
at the present.  We therefore speculate about future developments
involving the MEM, without regard to (current) computational
limitations.

\begin{itemize}
\item \textbf{NLO/ Parton Showers--}  The extension of the MEM to take
  into account additional radiation and/or other NLO corrections has
  already been considered~\cite{memhzz nlo, MEM HiggsNLO, extra rad}.
  Work has also gone into extending the MEM to include parton
  showers~\cite{shower deconstruction}.
Such approaches, potentially extended beyond NLO, will leverage the
more complete higher order calculations of the future.

\item \textbf{Jet Substructure--}  As outlined in Eq.~\ref{eq:mem1},
  the MEM is only using the four momentum of a jet.  However,
  other properties of the jet,
  e.g. substructure~\cite{substructure} could give additional
  information about the hard process parton with which a jet should be
  associated, thereby increasing the sensitivity of the analysis.
\item \textbf{Underlying Event, Hadronization, Etc.--} Additional
  information about hadronization, the underlying event, and other
  topics could also be used to calculate a more complete likelihood
  for events in hadron colliders.
\item \textbf{Detector Resolution--} Ultimately, instead of using a
  single transfer function for a detector, a separate transfer
  function could be utilized for each detector element involved in the
  reconstruction of an event.  Ideally the time dependence of the
  element response, as well as its correlations with other detector
  elements would be included.
\end{itemize}

\textbf{Conclusions--}
The future developments of the MEM described above will require heroic
efforts of theorists and experimentalists\footnote{An annual workshop
  to discuss MEM developments was initiated this year~\cite{workshop}}.  The result will be a
significant increase in the ability of future colliders to perform
precision measurements and searches.

\textbf{Acknowledgments--} 
MP is supported by the CERN-Korea fellowship through the National Research
Foundation of Korea.  Work supported in part by U.S. Department of
Energy Grants DE-FG02-97ER41029.  Fermilab is operated by the Fermi
Research Alliance under contract DE-AC02-07CH11359 with the
U.S. Department of Energy.


\begin{thebibliography}{99}
\bibitem{Bhat:2010zz} 
  P.~C.~Bhat,
  ``Multivariate Analysis Methods in Particle Physics,''
  Ann.\ Rev.\ Nucl.\ Part.\ Sci.\  {\bf 61}, 281 (2011).

\bibitem{matrix}
  K.~Kondo,
  J.\ Phys.\ Soc.\ Jap.\  {\bf 57}, 4126 (1988) and
  {\bf 60}, 836 (1991);
  R.~H.~Dalitz and G.~R.~Goldstein,
  Phys.\ Rev.\  D {\bf 45}, 1531 (1992).

\bibitem{Neyman:Pearson}
J.~Neyman and E.~S.~Pearson,
Phil.\ Trans.\ R.\ Soc.\ Lond.\ A {\bf 231} no. 694-706, 289-337 (1933).

\bibitem{memtev}
  B.~Abbott {\it et al.}  [D\O\ Collaboration],
  Phys.\ Rev.\  D {\bf 60}, 052001 (1999);
  J.~C.~Estrada Vigil,
  FERMILAB-THESIS-2001-07;
  M.~F.~Canelli,
  UMI-31-14921;
  V.~M.~Abazov {\it et al.}  [D\O\ Collaboration],
  Nature {\bf 429}, 638 (2004);
  A.~Abulencia {\it et al.}  [CDF Collaboration],
  Phys.\ Rev.\  D {\bf 74}, 032009 (2006);
  A.~Abulencia {\it et al.}  [CDF Collaboration],
  Phys.\ Rev.\  D {\bf 75}, 031105 (2007);
  V.~M.~Abazov {\it et al.}  [D\O\ Collaboration],
  Phys.\ Rev.\  D {\bf 78}, 012005 (2008);
  T.~Aaltonen {\it et al.}  [CDF Collaboration],
  Phys.\ Rev.\ Lett.\  {\bf 101}, 252001 (2008);
  F.~Fiedler, A.~Grohsjean, P.~Haefner and P.~Schieferdecker,
  Nucl.\ Instrum.\ Meth.\  A {\bf 624}, 203 (2010);
  F.~Fiedler, A.~Grohsjean, P.~Haefner and P.~Schieferdecker,
  Nucl.\ Instrum.\ Meth.\  A {\bf 624}, 203 (2010);
  T.~Aaltonen {\it et al.}  [CDF Collaboration],
  CDF/PHYS/TOP/PUBLIC/10191 (2010).

\bibitem{B physics}
  I.~Dunietz, H.~R.~Quinn, A.~Snyder, W.~Toki and H.~J.~Lipkin,
  Phys.\ Rev.\ D {\bf 43}, 2193 (1991);
  G.~Kramer and W.~F.~Palmer,
  Phys.\ Rev.\ D {\bf 45}, 193 (1992);
  A.~V.~Gritsan and J.~G.~Smith,
  ``Polarization in B Decays,'' in
  J.~Beringer {\it et al.}  [Particle Data Group Collaboration],
  Phys.\ Rev.\ D {\bf 86}, 010001 (2012).


  \bibitem{memzz}
  A.~De Rujula, J.~Lykken, M.~Pierini, C.~Rogan and M.~Spiropulu,
  Phys.\ Rev.\ D {\bf 82}, 013003 (2010)
  [arXiv:1001.5300 [hep-ph]];
  J.~S.~Gainer, K.~Kumar, I.~Low and R.~Vega-Morales,
  JHEP {\bf 1111}, 027 (2011)
  [arXiv:1108.2274 [hep-ph]];
  D.~Stolarski and R.~Vega-Morales,
  Phys.\ Rev.\ D {\bf 86}, 117504 (2012)
  [arXiv:1208.4840 [hep-ph]];
  T.~Modak, D.~Sahoo, R.~Sinha and H.~-Y.~Cheng,
  arXiv:1301.5404 [hep-ph].

\bibitem{memhzz nlo}
  J.~M.~Campbell, W.~T.~Giele and C.~Williams,
  arXiv:1205.3434 [hep-ph];
  J.~M.~Campbell, W.~T.~Giele and C.~Williams,
  JHEP {\bf 1211}, 043 (2012)
  [arXiv:1204.4424 [hep-ph]].

\bibitem{jhu mem}
  Y.~Gao, A.~V.~Gritsan, Z.~Guo, K.~Melnikov, M.~Schulze and N.~V.~Tran,
  Phys.\ Rev.\ D {\bf 81}, 075022 (2010)
  [arXiv:1001.3396 [hep-ph]];
  S.~Bolognesi, Y.~Gao, A.~V.~Gritsan, K.~Melnikov, M.~Schulze, N.~V.~Tran and A.~Whitbeck,
  Phys.\ Rev.\ D {\bf 86}, 095031 (2012)
  [arXiv:1208.4018 [hep-ph]].

\bibitem{mekd}
  P.~Avery, D.~Bourilkov, M.~Chen, T.~Cheng, A.~Drozdetskiy, J.~S.~Gainer, A.~Korytov and K.~T.~Matchev {\it et al.},
  Phys.\ Rev.\ D {\bf 87}, 055006 (2013)
  arXiv:1210.0896 [hep-ph].

\bibitem{geolocating}
  J.~S.~Gainer, J.~Lykken, K.~T.~Matchev, S.~Mrenna and M.~Park,
  arXiv:1304.4936 [hep-ph].

\bibitem{MELA CMS}
  S.~Chatrchyan {\it et al.}  [CMS Collaboration],
  JHEP {\bf 1204}, 036 (2012)
  [arXiv:1202.1416 [hep-ex]];
  S.~Chatrchyan {\it et al.}  [CMS Collaboration],
  Phys.\ Lett.\ B {\bf 716}, 30 (2012)
  [arXiv:1207.7235 [hep-ex]].

\bibitem{MELA MEKD CMS}
  S.~Chatrchyan {\it et al.}  [CMS Collaboration],
  Phys.\ Rev.\ Lett.\  {\bf 110}, 081803 (2013)
  [arXiv:1212.6639 [hep-ex]];
  [CMS Collaboration],
  CMS-PAS-HIG-12-041;
  [CMS Collaboration],
  CMS-PAS-HIG-13-002.

\bibitem{ATLAS}
  G.~Aad {\it et al.}  [ATLAS Collaboration],
  arXiv:1307.1432 [hep-ex].

\bibitem{mem higgs}
  K.~Cranmer and T.~Plehn,
  Eur.\ Phys.\ J.\ C {\bf 51}, 415 (2007)
  [hep-ph/0605268];
S.-C.~Hsu {\it et al.}  [CDF Collaboration], CDF note 8774 (2007);
  T.~Aaltonen {\it et al.}  [CDF Collaboration],
  Phys.\ Rev.\ D {\bf 80}, 071101 (2009);
J.~Therhaag, Diplom thesis, University of Bonn, BONN-IB-2009-03 (2009);
 J.~S.~Gainer, W.-Y.~Keung, I.~Low and P.~Schwaller,
  Phys.\ Rev.\ D {\bf 86}, 033010 (2012);
 J.~R.~Andersen, C.~Englert and M.~Spannowsky,
  arXiv:1211.3011 [hep-ph];
  J.~R.~Andersen, C.~Englert and M.~Spannowsky,
  Phys.\ Rev.\ D {\bf 87}, 015019 (2013)
  [arXiv:1211.3011 [hep-ph]];
  A.~Freitas and J.~S.~Gainer,
  arXiv:1212.3598 [hep-ph];
  P.~Artoisenet, P.~de Aquino, F.~Maltoni and O.~Mattelaer,
  arXiv:1304.6414 [hep-ph];
  P.~Artoisenet, P.~de Aquino, F.~Demartin, R.~Frederix, S.~Frixione, F.~Maltoni, M.~K.~Mandal and P.~Mathews {\it et al.},
  arXiv:1306.6464 [hep-ph].


\bibitem{MEM HiggsNLO}
  J.~M.~Campbell, R.~K.~Ellis, W.~T.~Giele and C.~Williams,
  Phys.\ Rev.\ D {\bf 87}, 073005 (2013)
  [arXiv:1301.7086 [hep-ph]].

\bibitem{MEM BSM}
  J.~Alwall, A.~Freitas and O.~Mattelaer,
  AIP Conf.\ Proc.\  {\bf 1200}, 442 (2010);
  C.-Y.~Chen and A.~Freitas,
  JHEP {\bf 1102}, 002 (2011);
  O.~Gedalia, G.~Isidori, F.~Maltoni, G.~Perez, M.~Selvaggi and Y.~Soreq,
  arXiv:1212.4611 [hep-ph].

\bibitem{MadWeight}
  P.~Artoisenet, V.~Lema\^itre, F.~Maltoni and O.~Mattelaer,
  JHEP {\bf 1012}, 068 (2010).

\bibitem{Moore}
G.~E.~Moore,
Electronics, {\bf 38}, Number 8, 4 (1965).

\bibitem{GPU}
\verb*#https://en.wikipedia.org/wiki/Graphics_processing_unit#

\bibitem{extra rad}
  J.~Alwall, A.~Freitas and O.~Mattelaer,
  Phys.\ Rev.\ D {\bf 83}, 074010 (2011).

\bibitem{shower deconstruction}
  D.~E.~Soper and M.~Spannowsky,
  Phys.\ Rev.\ D {\bf 84}, 074002 (2011)
  [arXiv:1102.3480 [hep-ph]];
  D.~E.~Soper and M.~Spannowsky,
  arXiv:1211.3140 [hep-ph].

\bibitem{substructure}
  J.~Shelton,
  ``TASI Lectures on Jet Substructure,''
  arXiv:1302.0260 [hep-ph].

\bibitem{workshop}
\verb^https://agenda.irmp.ucl.ac.be/^\\
\verb^conferenceDisplay.py?confId=1502^

\end{thebibliography}
\end{document}